\documentstyle[11pt,psfig,twoside]{article}
\input epsf.sty
\begin{document}

\newcommand{\Abstract}[2]{{\footnotesize\begin{center}ABSTRACT\end{center}
\vspace{1mm}\par#1\par
\noindent
{~}{\it #2}}}

\newcommand{\TabCap}[2]{\begin{center}\parbox[t]{#1}{\begin{center}
 \small {\spaceskip 2pt plus 1pt minus 1pt T a b l e}
 \refstepcounter{table}\thetable \\[2mm]
 \footnotesize #2 \end{center}}\end{center}}

\newcommand{\TableSep}[2]{\begin{table}[p]\vspace{#1}
\TabCap{#2}\end{table}}

\newcommand{\FigCap}[1]{\footnotesize\par\noindent Fig.\ %
 \refstepcounter{figure}\thefigure. #1\par}

\newcommand{\TableFont}{\footnotesize}
\newcommand{\TableFontIt}{\ttit}
\newcommand{\SetTableFont}[1]{\renewcommand{\TableFont}{#1}}

\newcommand{\MakeTable}[4]{\begin{table}[htb]\TabCap{#2}{#3}
 \begin{center} \TableFont \begin{tabular}{#1} #4 
 \end{tabular}\end{center}\end{table}}

\newcommand{\MakeTableSep}[4]{\begin{table}[p]\TabCap{#2}{#3}
 \begin{center} \TableFont \begin{tabular}{#1} #4 
 \end{tabular}\end{center}\end{table}}

\newenvironment{references}%
{
\footnotesize \frenchspacing
\renewcommand{\thesection}{}
\renewcommand{\in}{{\rm in }}
\renewcommand{\AA}{Astron.\ Astrophys.}
\newcommand{\AAS}{Astron.~Astrophys.~Suppl.~Ser.}
\newcommand{\ApJ}{Astrophys.\ J.}
\newcommand{\ApJS}{Astrophys.\ J.~Suppl.~Ser.}
\newcommand{\ApJL}{Astrophys.\ J.~Letters}
\newcommand{\AJ}{Astron.\ J.}
\newcommand{\IBVS}{IBVS}
\newcommand{\PASP}{P.A.S.P.}
\newcommand{\Acta}{Acta Astron.}
\newcommand{\MNRAS}{MNRAS}
\renewcommand{\and}{{\rm and }}
\section{{\rm REFERENCES}}
\sloppy \hyphenpenalty10000
\begin{list}{}{\leftmargin1cm\listparindent-1cm
\itemindent\listparindent\parsep0pt\itemsep0pt}}%
{\end{list}\vspace{2mm}}

\def\TYLDA{~}
\newlength{\DW}
\settowidth{\DW}{0}
\newcommand{\dw}{\hspace{\DW}}

\newcommand{\refitem}[5]{\item[]{#1} #2
\def\REFARG{#3}\ifx\REFARG\TYLDA\else, {\it#3}\fi
\def\REFARG{#4}\ifx\REFARG\TYLDA\else, {\bf#4}\fi
\def\REFARG{#5}\ifx\REFARG\TYLDA\else, {#5}\fi.}

\newcommand{\Section}[1]{\section{#1}}
\newcommand{\Subsection}[1]{\subsection{#1}}
\newcommand{\Acknow}[1]{\par\vspace{5mm}{\bf Acknowledgements.} #1}
\pagestyle{myheadings}

\def\thefootnote{\fnsymbol{footnote}}
\begin{center}
{\Large\bf Nonradial modes in RR Lyrae stars\\}
\vskip3pt
{\bf W.~A.~~D~z~i~e~m~b~o~w~s~k~i$^1$~~ and~~
S.~~C~a~s~s~i~s~i$^2$}
\vskip3mm
{$^1$Warsaw University Observatory, Al.~Ujazdowskie~4, 00-478~Warszawa, Poland\\
e-mail: wd@astrouw.edu.pl\\
$^2$ Osservatorio Astronomico Collurania, I-64100, Teramo, Italy\\
e-mail: cassisi@astrte.te.astro.it}
\vskip5mm
\end{center}

\Abstract{We present a survey of nonradial mode properties in evolutionary 
sequences of RR Lyrae star models. Attention is focused on the  
modes that may be driven by the opacity mechanism and on 
those that may be excited as a consequence of the 1:1 resonance 
with the radial pulsation.

Qualitatively, all the models share the same properties of the 
nonradial modes. At the quantitative level,  the   
properties are to a large extent determined by the radial mode periods. There
is only weak dependence on the star metallicity and 
no apparent dependence on the evolutionary status, that is on the helium 
exhaustion in the convective core.

In the whole range of RRab and RRc star parameters we find unstable nonradial
modes driven by the opacity mechanism. An instability of radial pulsation 
to a resonant excitation of nonradial oscillations is also a common 
phenomenon  in 
both types. We discuss a possible role of nonradial modes in 
amplitude modulation observed in certain RR Lyrae stars.}

\Section{Introduction}

Whether or not nonradial modes play a role in RR Lyrae pulsation 
has been a matter of speculation for some time. Recently, Olech et al. 
(1999) presented first circumstantial evidence for nonradial modes presence 
in three RRc 
variables in M55. The evidence was based on the power spectra which revealed 
presence of additional modes whose frequencies could not be attributed to 
radial modes.
Similar power spectra were subsequently found in several other
RRc anc RRab stars ( Olech et al. 1999b, Kovacs et al. 1999, 
Moskalik, P. 1999).

Earlier (Kovacs, 1993; Van Hoolst and Waelkens, 1995) 
proposed the 1:1 resonant excitation of nonradial modes in a radially
pulsating star 
as an explanation of the Blazkho type modulation. Manifestation 
of this effect in periodograms is an occurrence of the equally-spaced 
side peaks around the main frequency. Calculations provided by Van
Hoolst et al (1998) confirmed the plausibility of this idea.  
These authors studied stability of radial pulsation with use of the third 
amplitude equation formalism. They found that there is a high probability 
of a resonant excitation of a low $\ell$-degree mode. However, their
calculations were done only one stellar model. In this paper we apply
the same formalism, with one additional simplification, to investigate stability
of radial pulsation in a large sets of RR Lyrae star models.  
An outline of the formalism and the results are given 
in section 4.2.

Still earlier (Dziembowski, 1977) instability of 
static models of RR Lyrae stars to certain nonradial modes has 
been demonstrated.  The driving effect is the same as for radial 
pulsation. A linear instability, however, 
is not a sufficient condition for excitation. 
Nonlinear calculations 
are required to determine the ultimate outcome of the linear instability.
Because of enormous numerical complexity, for 
the nonradial modes such calculations have never been done.
The instability could be saturated with excitation of a single mode,
which seems the most common situation among RR Lyrae stars. However, 
for instance among $\delta$ Sct stars typical is excitation of 
many modes. We do not understand why is it so.
 
In section 4.1 of this paper we present a survey 
of the unstable nonradial modes in RR Lyrae stars and we discuss 
potential identifications of the modes detected by Olech et al. (1999).

\begin{figure}
\begin{center}
\epsfxsize=0.99\hsize
\epsffile{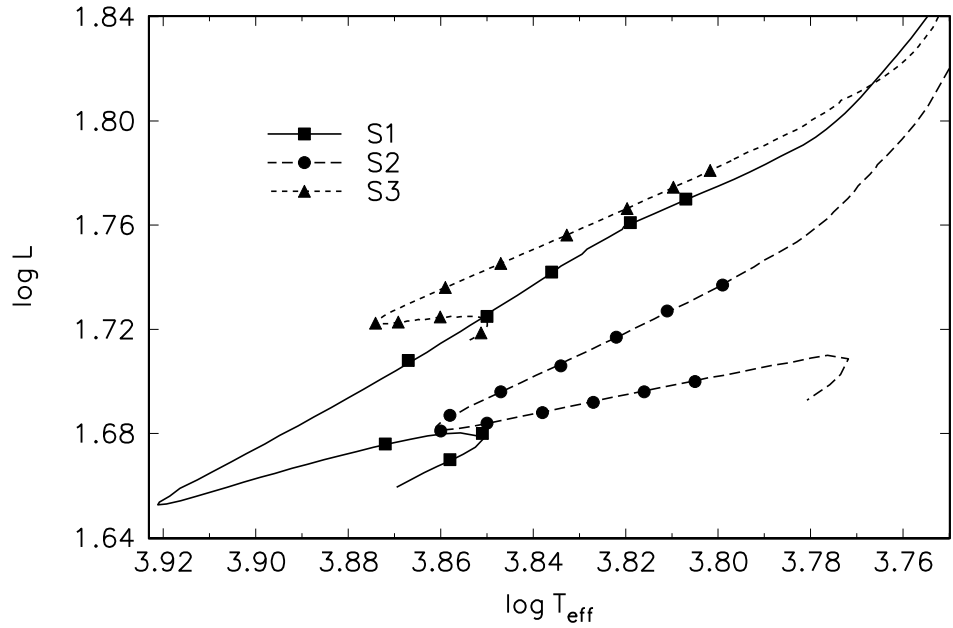}
\end{center}
\caption{Horizontal-branch evolutionary sequences.  
In S1 and S2 sequences the ZAHB composition is characterized by 
$Z=0.001, Y_{HB}=0.243$ and that in S3 by $Z=0.0002, Y_{HB}=0.24$. 
Stellar masses are 0.65, 0.67 and $0.74 M_\odot$, respectively, in S1, S2, S3.
The symbols mark the models selected for the pulsational analysis.}  
\end{figure}

\Section{Evolutionary models}

All the stellar models adopted in the present investigation, have been
computed adopting the latest version of the FRANEC evolutionary code, which
includes several upgrades of the input physics. Major improvements are the
opacity tables for the stellar interiors as given by Rogers \& Iglesias
(1992) and low-temperature molecular opacities for outer stellar
layers by Alexander \& Ferguson (1994). Both high- and low-temperature
opacity tables have been computed by adopting the Grevesse (1991) solar
chemical mixture. The equation of state is the OPAL one (Rogers et al. 1996),
implemented in the temperature-density region not covered by OPAL, with the
equation of state of Straniero (1988), plus a Saha Eos in the outer stellar
layers (see Cassisi et al. 1998, 1999, for more details).
As for the calibration of the superadiabatic envelope convection, the mixing
length calibration provided by Salaris \& Cassisi (1996) has been adopted.

For the present work, we have computed Horizontal-Branch models for two 
different assumptions on the heavy element abundance: Z=0.0002 and 0.001
namely. In both cases, an initial Helium abundance equal to Y=0.23 has been
adopted. All the HB models have as Red Giant Branch progenitor a structure
with mass equal to $0.8M_\odot$. This means that when computing the Zero Age
Horizontal Branch (ZAHB) models we have accounted for the evolutionary values 
for the size of the He core mass and the surface He abundance
($Y_{HB}$) at the He ignition corresponding to a $0.8M_\odot$
progenitor as provided by our own evolutionary computations for the
previous H-burning phases.

In Fig. 1, we show the selected 
evolutionary tracks in the H-R diagram. The symbols along 
each track indicate the models 
adopted for the following pulsational analysis.

\Section{Linear nonadiabatic calculations}

Oscillation properties of the selected models were studied with the 
method developed by one of us (Dziembowski, 1977). Its recent updated 
description may be found  in Van Hoolst et al. (1998). 
For nonradial modes the equation of 
nonadiabatic oscillations are solved numerically in the envelope and 
matched to the asymptotic solution for g-modes, which is valid in 
the deep interior of RR Lyrae stars. The reason is that 
beneath the matching point the Brunt-V\"ais\"al\"a 
frequency is much larger than the oscillation frequencies.
The Cowling approximation is assumed, which is well justified for 
the modes  considered. 
The weakest point in the adopted method is the one related to the
treatment of convective transport, whose Lagrangian perturbation is 
simply ignored. This is certainly a poor approximation but it is not 
essential for the main aim of this work  because effects of convection 
on radial and nonradial modes are nearly the same..  

As an introductory example we plot in Fig. 2 the growth rates,
 $\gamma=\Im(\omega)$, and 
frequencies, $f=\Re(\omega)/2\pi$ for modes at the  selected degrees $\ell$ for one of the 
models we chose for the pulsation analysis. The temporal dependence 
of oscillations is assumed in the form $\exp(-{\rm i}\omega t)$.
Effects of rotation has been ignored. Thus, each point represents 
$2\ell +1$ normal modes.

\begin{figure}
\begin{center}
\epsfxsize=0.99\hsize
\epsffile{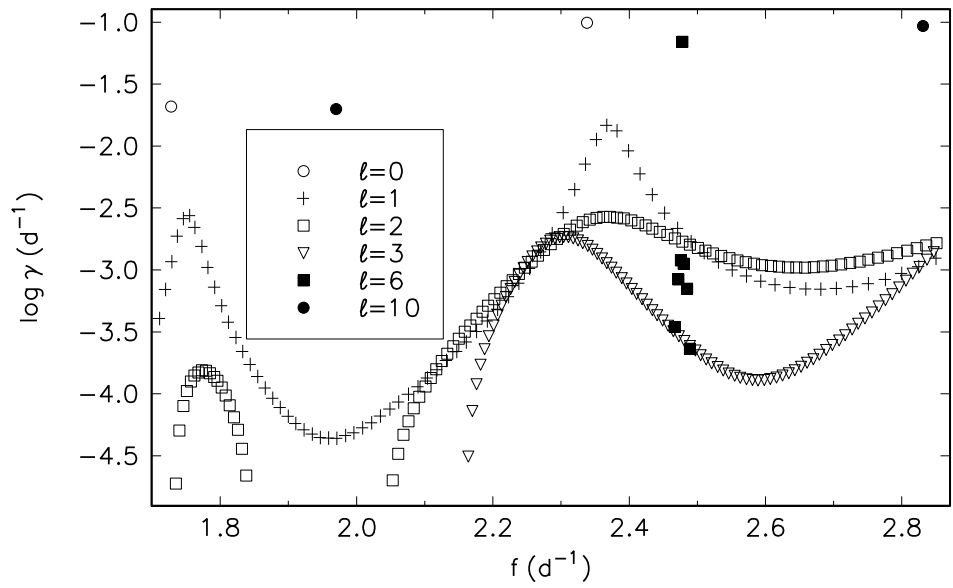}
\end{center}
\caption{Growth rates and frequencies for modes of indicated spherical 
harmonic degrees in a selected model from S3 sequence. 
The model is characterized by the following parameters
$\log T_{\rm eff}=3.822$, $\log(L/L_{\odot})=1.717$, $Y_c=0.17$}   
\end{figure}

Let us note two types of unstable modes. There are isolated rapidly 
unstable modes with the growth rates $\gamma > 0.01$ d$^{-1}$ and 
sequences of modes with much lower $\gamma$'s. In the former group we find 
radial modes and modes with $\ell$= 6 and 10 which belong to the 
class of strongly trapped unstable (STU) modes defined by Van Hoolst 
et al.(1998).
These modes have no counterpart in the adiabatic approximation. 
In the interior the eigenfunction of such modes are 
-- to good approximation -- described as inward propagating internal 
gravity waves with exponentially decreasing amplitude. 

The growth rate behavior in the sequences 
of low degree modes reflect the trapping properties of the acoustic cavity.
Still at $\ell=1$ even for the best trapped modes more than 80 percent 
of the kinetic energy is contributed by the g-mode propagation. The trapping 
effect is weaker at $\ell=2$ and 3 but then it begins to increase. 
Note the sharp peak of $\gamma$ in the $\ell=6$ sequence near the 
first overtone frequency.
The STU modes occur always between the two best trapped ordinary modes.

For the occurrence of STU mode a sufficient trapping in the evanescent 
zone separating p- and g-mode propagation zone is needed. 
In our selected mode the STU fundamental modes appear at $\ell=8$. 
With increasing $\ell$ they tend to Kelvin (f or surface) modes. 
The instability continues well above $\ell=100$. We hesitate to give 
the upper limit because of the increasing uncertainty due to our crude 
treatment of convection. Near the first overtone the STU modes begin at 
$\ell=5$ and end at $\ell=15$. 

\Section{Survey of pulsational properties}

The H-R positions of stellar models selected for this survey were 
shown in figure 1. The models cover various stages of the central 
helium burning. This can be seen in figure 3, where central helium 
content is plotted as function of the effective temperature. In the same 
figure, the periods of the first two radial modes are plotted. Solid symbols 
are used to denote linearly unstable modes. Second overtone is also 
unstable in some of our models. However, because three is no observational
evidence for second overtone excitation in RR Lyrae stars, in present 
survey we consider only modes in the vicinity of the first two radial modes.

\begin{figure}
\begin{center}
\epsfxsize=0.99\hsize
\epsffile{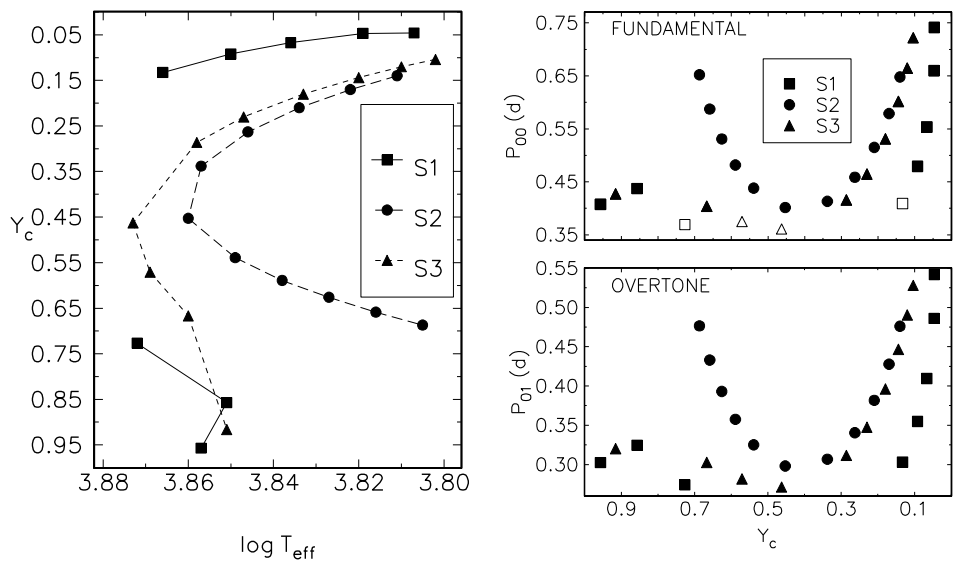}
\end{center}
\caption{Central helium content ($Y_c$) in selected models is plotted
as function of their effective temperature in the left panel. 
In the right two panels periods of the radial fundamental and first
overtone modes are given for the selected models. The empty symbols are used 
if the mode is stable}   
\end{figure}

\Subsection{Opacity-driven modes}

The general property of all models considered is that the trapping effect
is weak in the $\ell=2-4$ range. In the vicinity of 
an unstable fundamental radial there are always unstable $\ell=1$ modes.
In some models, like the one used in Fig. 2, there is a frequency range 
where the $\ell=2$ modes are unstable as well but with much lower growth 
rates. Rapid instability occurs only for the STU modes, which began in 
most of the models at $\ell=8$. 
The trapping pattern near first overtone is similar to that near the 
fundamental mode. Again the most unstable are modes of the $\ell=1$ degree
and then STU modes which begin at $\ell=5$ or 6.
The main difference is instability at all low degrees.

In Figs. 4 and 5 we show, respectively for  
the  fundamental mode and first overtone ranges, 
the frequency distances to corresponding radial   
modes and the relative growth rates for most unstable 
$\ell=1$ modes and for the selected STU modes. 
The most unstable $\ell=1$ modes as well as all the STU modes 
have always higher frequencies than the corresponding radial modes. 
The two plotted parameters vary in rather narrow ranges and their values 
are determined by the radial mode periods. The dependence on the 
abundance ($Z$) is most easily seen in the distances of the STU modes.
The dependence on the evolutionary  status ($Y_c$), for which  
one should consult Fig. 3, is not recognizable.

\begin{figure}
\begin{center}
\epsfxsize=0.99\hsize
\epsffile{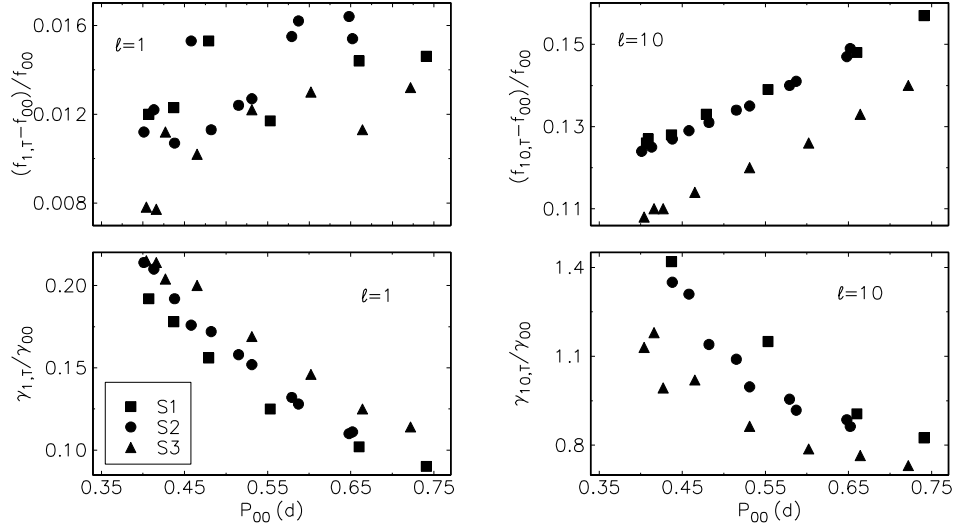}
\end{center}
\caption{The relative frequency separation between most trapped 
(subscript "T") nonradial 
modes and the fundamental radial mode (top). The relative (radial to nonradial)
growth rates (bottom). The $\ell=1$ mode correspond to the local maximum 
of $\gamma$. The $\ell=10$ mode is one of STU modes.}   
\end{figure}

The growth rates of STU modes are similar to those of the radial 
modes.
If the instability is saturated by one of these relatively high degree 
modes the star would appear as a nonpulsating object. 
At $\ell=5-10$ the cancellation of the opposite sign contribution  
would reduce the disc-averaged light amplitude   
to at most few millimagnitude. There is no firm evidence for 
occurrence of nonpulsating stars in RR Lyrae stars in the RR Lyrae 
domain of the H-R. The linear theory does not yield us a hint why 
radial modes are so much preferred over the STU modes by stars.  

\begin{table}

\caption{Amplitude-modulated RRc stars in M55 (from Olech et al.1999)} 

\begin{tabular}{cccc}
\hline
  Star      & $P_p[d]$  &  $(f_p-f_s)/f_p$ & $A_s/S_p$ \\
\hline
V9 & 0.316 & -0.028 & 0.19\\
V10 & 0.332 & +0.004 & 0.44\\
V9 & 0.316 & -0.098 & 0.57\\
\hline
\end{tabular}
\label{Table 1}
\end{table}

The secondary peaks in the three amplitude-modulated 
RRc stars discovered by Olech et al.(1999) cannot 
be explained in terms STU mode excitations. In Table 1 we 
provide data on  the distances between the primary 
and secondary peaks and the relative V-amplitudes.

The secondary peak amplitudes are 
still too large and in two cases the frequencies are lower than those
of the main peaks. Interpretation in terms of   
the $\ell=1$ modes is more plausible though not free of difficulties.
Also in this case the secondary peaks position present certain problem.
However, the problem is not so essential because always we find unstable 
$\ell=1$ modes on both sides of the radial modes.  
Furthermore, we do not have arguments why radial modes should always 
have higher amplitudes.

\begin{figure}
\begin{center}
\epsfxsize=0.99\hsize
\epsffile{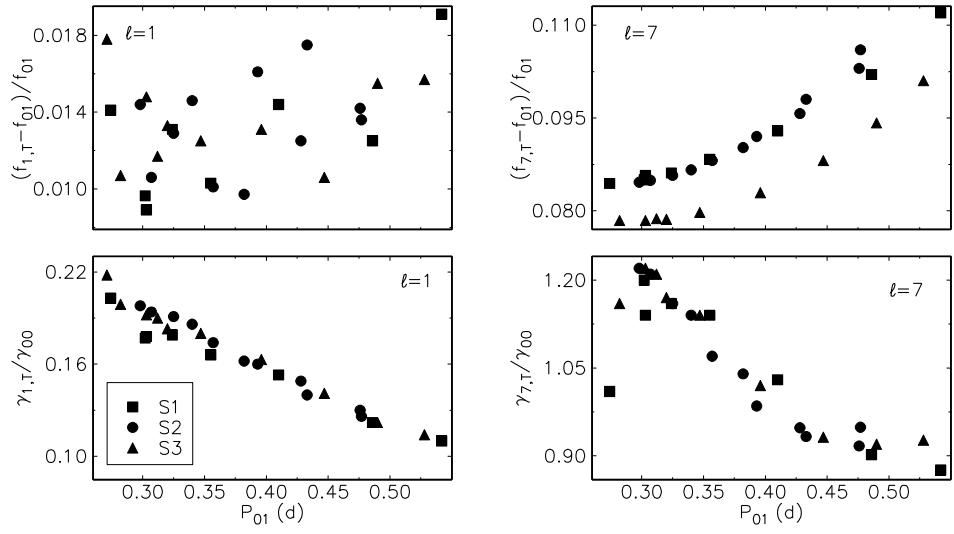}
\end{center}
\caption{
Similar to Fig.4 but for the first overtone vicinity. Here 
$\ell=7$ is the chosen STU mode.}   
\end{figure}

\Subsection{Resonant modes}

The criterion for the instability of radial pulsation to excitation
of a resonant nonradial mode may be written in the following form
\begin{equation}
\left({\delta R\over R}\right)^2> \sqrt{D^2 +\kappa^2\over C},
\end{equation} 
where $\delta R$ is the amplitude of radius variations in radial 
pulsation; $D$ denotes the frequency distance between the radial and
the nonradial mode; $\kappa$ is the damping rate 
of the nonradial mode in the limit cycle of the radial mode; 
and $C$ is the coupling coefficient.
The criterion is from Van Hoolst et al.(1998) only the notation is different.
 
\begin{figure}
\begin{center}
\epsfxsize=0.99\hsize
\epsffile{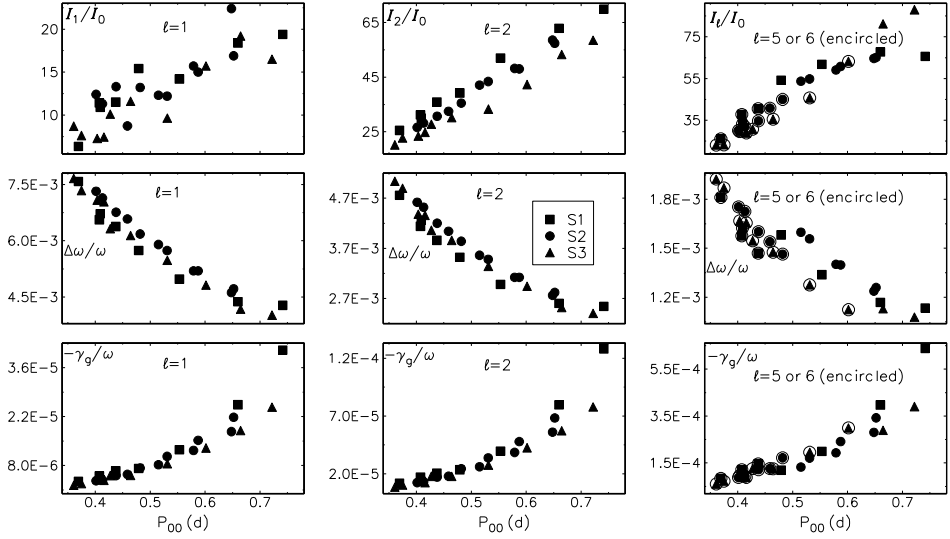}
\end{center}
\caption{ Properties of the resonant modes (closest to the fundamental 
radial mode).
From top to bottom, the relative moment of inertia, the relative 
frequency separation between consecutive modes, the fractional damping rate
are plotted against radial mode period for the fundamental 
mode vicinity. The parameters plotted are important for the resonant 
excitation of the nonradial modes. Modes with degrees $\ell=1$, and 5 or 6 
are most likely excited.}   
\end{figure}

Evaluation of the quantities  occurring in the r.h.s. of Eq.(1) requires, in 
principle, nonlinear calculations, which we have not done. However,  
for a crude evaluation  of the probability   
that a radial mode of specified amplitude is unstable to 
parametric excitation of a nonradial mode, we   
need only linear mode characteristic provided in Figs. 6 and 7 
and certain coupling coefficients. These coefficients were  
evaluated by Van Hoolst et al.(1998) for the stellar model they 
selected. Here we rely on a simple scaling of their numbers which 
we will explain below.

Let ${\cal P}_\ell(A)$ denotes the excitation probability 
at the radial mode amplitude $A=\delta R/R$. Then, if effect 
of rotation are ignored, we have 
\begin{equation}
{\cal P}_\ell(A) =\left\{ \begin{array}{ll}
0 & \mbox{if $A^4C^2\le\kappa^2$}\\
{\rm Min}\left(1,{1\over2}{\sqrt{A^4C^2-\kappa^2}\over\Delta\omega}\right) & 
\mbox{if  $A^4C^2>\kappa^2$} 
\end{array}
\right.
\end{equation}
where $\Delta\omega=\omega_{\ell,n-1}-\omega_{\ell,n}$ denotes the 
frequency distance between consecutive g-modes of degree $\ell$.
Note that $\Delta\omega/2$ is the maximum frequency distance between the 
radial and the nearest nonradial mode and that $\sqrt{A^4C^2-\kappa^2}$ is 
the distance at the onset of the instability.

The coupling coefficients, $C$, for various radial \& nonradial 
mode pairs in the model of RR Lyrae star was explicitly calculated by 
Van Hoolst et al. (1998) . From their numbers
we found an approximate relation 
\begin{equation}
 C_{k,\ell}=b_k{I_{0,k}\over I_\ell},
\end{equation} 
with
$$b_0=27\quad\mbox{ and }\quad b_1=172 d^{-1},$$
where $k$ denote radial mode order (here $k=0$ and 1 for the fundamental
and the first overtone, respectively), $I$ denotes mode inertia evaluated 
assuming the same amplitude at the surface. 
Our additional simplification consists in adopting the same $b_k$ values
for all our models.

Final simplification, which we adopted after Van Hoolst et al. (1998),
is the assumption that
$\kappa=-\gamma_g$,
where -$\gamma_g$ is the damping rate due to dissipation in the 
g-mode propagation zone. This seems well justified because we consider 
the situation when the opacity driven instability is saturated by the 
radial mode and the resonant 
nonradial modes have almost the same properties in the outer layers.
Consequently, there should be also an exact balance 
between the driving and damping also for the nonradial mode. 
Then -$\gamma_g$  is all what remains.

The joint probability of the instability is given
by 
\begin{equation}
{\cal P}(A)=1-\prod_{\ell}\left[1-{\cal P}_\ell(A)\right]. 
\end{equation}
For rotating stars we have to consider modes of different
azimuthal numbers and evaluate probabilities ${\cal P}_{\ell,m}$.
The effect increases probability of the resonant instability (Dziembowski
et al. 1988).

\begin{figure}
\begin{center}
\epsfxsize=0.99\hsize
\epsffile{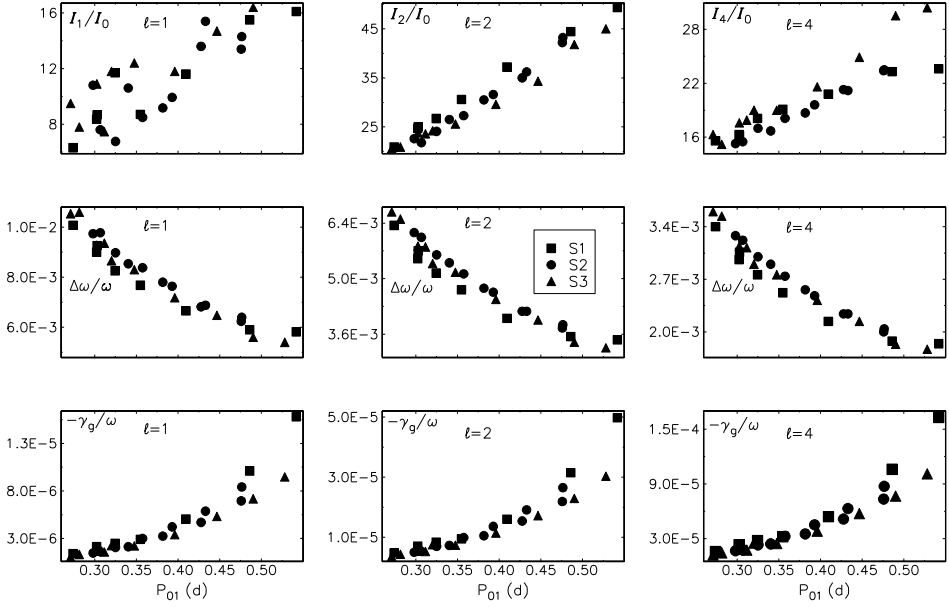}
\end{center}
\caption{Similar to Fig.6 but for the first overtone. 
In this case the $\ell=1$ and 4 are most likely excited.}   
\end{figure}

Very much like Van Hoolst et al. (1998) we find the maximum of probability
of the resonant excitation of an $\ell=1$ mode, both for the fundamental and 
the overtone radial modes and then an $\ell=5$ or 6 mode for the fundamental 
and an $\ell=4$ for the overtone. This is why we selected these $\ell$-values
in Figs. 6 and 7.  
In addition, there are data for $\ell=2$ modes. Excitation of these 
modes is less likely than the $\ell=1$ modes because of higher inertia.
The data on $I_2/I_0$ are important for 
evaluation of effects of rotation and magnetic 
field on radial pulsation. We will not discuss these effects here.

\begin{figure}
\begin{center}
\epsfxsize=0.99\hsize
\epsffile{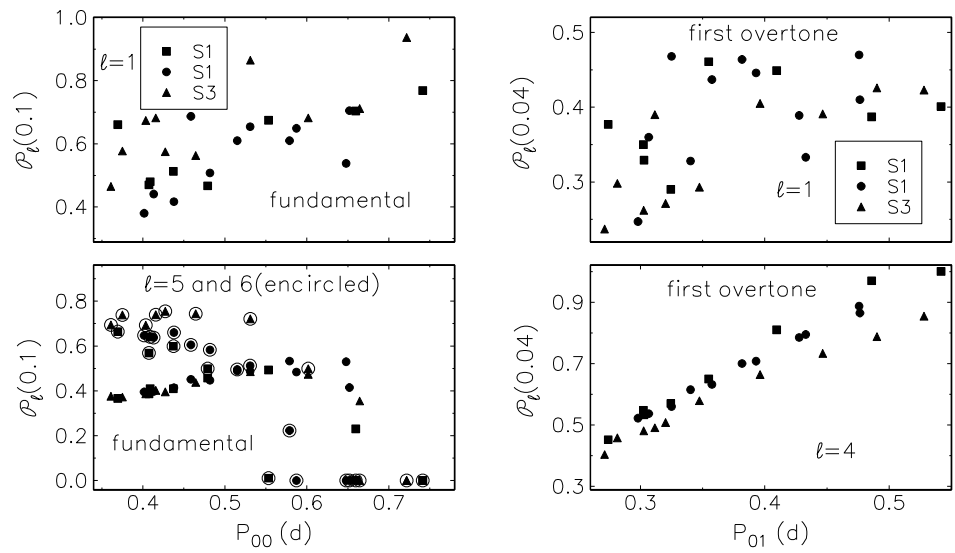}
\end{center}
\caption{Probability of instability of radial pulsation in the fundamental 
and first overtone mode relative to excitation of nonradial modes of indicated 
degrees.}   
\end{figure}

In Fig. 8 we present results of calculation of the excitation probabilities
for modes of selected degrees which yield the dominant contribution to the 
joint probability. We chose 
$A=0.075$ for the fundamental mode and  $A=0.025$ for the first overtone.
These values correspond to the mean amplitudes of radius variations   
in, respectively, RRab and RRc stars 
(e.g. Jones et al. 1988,Capricorni et al. 1989, Liu \& Janes, 1990, 
Jones et al. 1992). The lower probability of the first overtone instability is 
a direct consequence of the lower value of $A$.

The probability of the resonant instability in most cases increases
with pulsation period. The exception is the instability of the fundamental 
mode to higher degree modes. In this case damping in the 
g-mode propagation zone plays an important role. The value of $\kappa$ 
increases with $P_0$ and $\ell$. The increase
reduces the chances for the instability and ultimately
prevents it (see Eq. 2).
 
At typical amplitude of RRab the probability of excitation 
of $\ell=1$ mode is between 0.25 and 0.5. This is not so different 
from the incidence of Blazkho effect which is estimated to be between 
20\% and 30\%. The joint probability of the instability is always 
higher than 0.5 and close to 1 in most cases. However, excitation of
modes with $\ell>2$ may not lead to amplitude modulation.
The incidence of Blazkho effect amongst RRc stars is lower. 
Kovacs et al.(1999) who analyzed data on a large sample of RRc stars from LMC
found the effect in 1.4\% of the objects. Our analysis suggest lower chances 
for the first overtone instability than the fundamental mode but not in such
a disproportion.

\Section{Conclusions and discussion}

Our survey shows that  
all RR Lyrae star models share all qualitative properties of the nonradial
modes. There is always a large number of unstable low degree modes with 
frequencies close to unstable radial modes.
However, owing to higher mode inertia, for most of nonradial modes
the driving rates are much lower than those for radial modes. 
 The exceptions are the strongly trapped (STU) 
modes which begin with $\ell$ degrees 7 to 10 (depending on the model) at 
frequencies  somewhat above the fundamental radial mode and with $\ell=5$ 
or 6 with  at frequencies  somewhat above the radial mode overtones.
These modes are characterized  by growth rates similar to radial modes.
However, we argued that these modes are not likely candidates for 
identification of oscillation detected in some RR Lyrae stars. 
More likely candidates are the $\ell=1$ modes. Their driving rates are 
by nearly an order of magnitude lower than radial modes but it is well known 
thatthe growth rate is not necessarily a good 
predictor of the finite amplitude pulsation.

We found also that parameters which determine the chances of the excitation 
of nonradial radial modes through the 1:1 resonance do not vary much 
over the range of RR Lyrae stars parameters. According to our estimate 
the excitation has a high probability. In fact some nonradial modes should
be excited in majority of the RRab pulsators and in a significant fraction
of $(\>30\%)$ of RRc pulsators. The actual number should be greater because 
we ignored effect of rotation. Our crude estimate, which we did not detail 
here, shows that the effest is significant already at equatorial velocities of 
few km/s.

Why then the incidence of the anomalous behavior 
among RR Lyrae stars is relatively rare?   
We should stress, that a significant amplitude modulation is not 
automatically implied by the nonradial mode excitation.
If the nonlinear interaction between radial and nonradial modes 
leads to a steady  pulsation with constant amplitude then the presence 
of the nonradial mode will not be easy detectable. A Blazkho-type amplitude 
modulation may arise then only if the nonradial mode is not symmetric
about the rotation axis and it is of low degree. In this case the Blazkho 
period is equal to the rotation period.
Another possibility is a periodic limit cycle in which the amplitudes of the 
two modes vary intrinsically with the period determined by the nonlinear
interaction. 

The ulitmate answer regarding the presence of nonradial modes in RR Lyrae 
stars may be expected only from spectropy. A signature of such modes 
should be searched in the line-profile variations. Thus, 
high-resolution spectroscopic observations of amplitude
modulated RR Lyrare stars are encouraged.

\Acknow{The paper was supported by the KBN grant 2P03D00814.
S.C. warmly thanks for the hospitality during his stay at the
Copernicus Astronomical Center. We are grateful to Geza Kovacs for 
reading a preliminary version of this paper and a number 
of useful suggestions }

\end{document}